\documentclass{aastex}
\usepackage{emulateapj5}
\usepackage{epsf}
\shorttitle{AGN FEEDBACK ON THE ICM} \shortauthors{CAVALIERE, LAPI
\& MENCI}
\begin{document}
%%%%%%%%%%%%%%%%%%%%%%%%%%%%%%%%%%%%%%%%%%%%%%%%%%%%%%%%%%%%%%%%%%%%%%%%%
%%%%%%%%%%%%%%%%%%%%%%%%%%%%%%%%%%%%%%%%%%%%%%%%%%%%%%%%%%%%%%%%%%%%%%%%%
\title{AGN Feedback on the Intracluster Medium}
%%%%%%%%%%%%%%%%%%%%%%%%%%%%%%%%%%%%%%%%%%%%%%%%%%%%%%%%%%%%%%%%%%%%%%%%%
%%%%%%%%%%%%%%%%%%%%%%%%%%%%%%%%%%%%%%%%%%%%%%%%%%%%%%%%%%%%%%%%%%%%%%%%%
%
%%%%%%%%%%%%%%%%%%%%%%%%%%%%%%%%%%%%%%%%%%%%%%%%%%%%%%%%%%%%%%%%%%%%%%%%%
\author{A. Cavaliere and A. Lapi}
\affil{Astrofisica, Dipartimento di Fisica, Universit\'a `Tor
Vergata', Roma, I-00133}
\and
\author{N. Menci}
\affil{Osservatorio Astronomico di Roma, Monteporzio Catone,
I-00044}
%%%%%%%%%%%%%%%%%%%%%%%%%%%%%%%%%%%%%%%%%%%%%%%%%%%%%%%%%%%%%%%%%%%%%%%%%
%
\begin{abstract}
Galaxy groups are quite underluminous in X-rays compared to
clusters, so the intracluster medium has to be considerably
underdense in the former. We consider this to be due to
substantial energy fed back into the ICM when the baryons in the
member galaxies condense into stars ending up in SNe, or accrete
on to central supermassive black holes energizing AGNs. We compute
the outflow and the blowout effects driven by the AGNs and the
resulting, steep luminosity-temperature correlation $L_X-T$. We
compare this with the SN contribution and with the X-ray data; the
latter require the AGN energy to be coupled to the surrounding ICM
at fractional levels around $5\, 10^{-2}$. We link the $L_X-T$
behavior with the parallel effects of the AGN feedback on the gas
in the host galaxy; we find that these yield a correlation steep
up to $M_{\bullet}\propto \sigma^5$ between the galactic velocity
dispersions and the central BH masses.
\end{abstract}
\keywords{galaxies: clusters: general - quasars: general - shock
waves - X-rays: galaxies: clusters}
%
%%%%%%%%%%%%%%%%%%%%%%%%%%%%%%%%%%%%%%%%%%%%%%%%%%%%%%%%%%%%%%%%%%%%%%%%%
%%%%%%%%%%%%%%%%%%%%%%%%%%%%%%%%%%%%%%%%%%%%%%%%%%%%%%%%%%%%%%%%%%%%%%%%%
\section{Introduction}
%%%%%%%%%%%%%%%%%%%%%%%%%%%%%%%%%%%%%%%%%%%%%%%%%%%%%%%%%%%%%%%%%%%%%%%%%
%%%%%%%%%%%%%%%%%%%%%%%%%%%%%%%%%%%%%%%%%%%%%%%%%%%%%%%%%%%%%%%%%%%%%%%%%
%
Groups and clusters of galaxies shine in X-rays due to the thermal
bremsstrahlung radiation emitted by the hot intracluster medium
(ICM) they contain.

But the poorer groups are found to be progressively underluminous,
so as to lie substantially below the simple scaling $L_X\propto
T_v^2$. For the latter to hold, in the luminosity $L_X\propto
n^2\, R^3\, T^{1/2}$ the ICM number density $n$ would have to be
proportional to the gravitationally dominant DM mass density
$\rho$. This constraint adds to the temperature $T$ being close
the virial value $T_v$ and the size $R$ scaling as the virial
radius $R_v\propto T_v^{1/2}\, \rho^{-1/2}$.

In fact, the observed $L_X-T$ correlation has a shape more like
$L_X\propto T_v^3$ for richness $1$ clusters (Kaiser 1991), and in
moving toward poor groups it bends further down to $L_X\propto
T_v^{5}$ or steeper (Ponman, Cannon, \& Navarro 1999; see also
\S~3). So in groups the ICM is considerably underdense relative to
the cluster values $n\approx 10^{-1}\, \rho/m_p$.

Correspondingly, the central ICM entropy $S/k = \ln kT/n^{2/3}$
deviates upwards from the simple scaling $e^S\propto T_v$, to
attain in poor groups the `floor' value $kT/n^{2/3}\approx 140$
keV cm$^2$ (Lloyd-Davies, Ponman \& Cannon 2000). This requires a
density deficit associated with increased or constant $T/T_v$.
Such a non-adiabatic behavior may be traced back to energy losses
from, or additions to the ICM.

Here we focus on the `heating' that arises as the baryons in the
member galaxies condense into stars followed by SNe, or accrete
onto a supermassive black hole (BH) kindling an AGN. Two issues
stand in the way: (i) any extra energy from sources has to compete
with the huge thermal value $E\approx 10^{61} \,
(kT_v/\mathrm{keV})^{5/2}$ erg for the ICM in equilibrium; (ii)
the extra energy that can be coupled to it is still poorly known.
We discuss these issues, and derive two observables that bound or
probe at group and at galactic scales the amount of energy
coupled.
%
%%%%%%%%%%%%%%%%%%%%%%%%%%%%%%%%%%%%%%%%%%%%%%%%%%%%%%%%%%%%%%%%%%%%%%%%%
%%%%%%%%%%%%%%%%%%%%%%%%%%%%%%%%%%%%%%%%%%%%%%%%%%%%%%%%%%%%%%%%%%%%%%%%%
\section{Feedback from SNe}
%%%%%%%%%%%%%%%%%%%%%%%%%%%%%%%%%%%%%%%%%%%%%%%%%%%%%%%%%%%%%%%%%%%%%%%%%
%%%%%%%%%%%%%%%%%%%%%%%%%%%%%%%%%%%%%%%%%%%%%%%%%%%%%%%%%%%%%%%%%%%%%%%%%
%
Obvious first candidates for energy discharges into the ICM are
the SN explosions following star formation in the member galaxies
of groups and clusters. Prompt, type II SNe canonically release
$10^{51}$ ergs; these are effectively coupled to the gas when SN
remnants propagate cooperatively over galactic scales to drive
galactic winds (Ostriker \& McKee 1988; Wang et al. 2001; Pettini
et al. 2001; Heckman 2002). With a coupling around $1/2$, the
energy input (including winds from hot stars) comes to $\Delta
E\la 3\, 10^{48}$ erg per solar mass condensed into stars. In a
fiducial group with $kT_v\approx 1$ keV, virial mass $M_v\approx
5\, 10^{13}\, M_{\odot}$ and stellar mass around $3\, 10^{12}\,
M_{\odot}$, this would raise by $k\Delta T\approx 0.3$ keV the
temperature of the entire ICM. The outcome looks like a modest
$\Delta E/E= \Delta T/T_v\approx 1/3$.

Actually, SNe make optimal use of their energy in that they
produce {\it hierarchical preheating} of the ICM, while a group
and its ICM are built up hierarchically through merging events
with a range of partners. About half the final DM mass $M_v$ is
contributed to the main progenitor by smaller partners with masses
within the window $M_v/3$ to $M_v/20$, corresponding to
temperatures from $0.6$ down to $0.15\, T_v$ (Lacey \& Cole 1993;
Menci \& Cavaliere 2000).

Smaller lumps in the window have shallower gravitational wells and
produce more star-related energy on scales closer to the dynamical
time $t_d$; so they are more effective in heating/ejecting their
gas share. During each subsequent step of the hierarchy such gas
preheated {\it externally} will less easily flow into the main
progenitor's well. Thus the effects propagate up the hierarchy,
and lower ICM densities are induced in all structures up to poor
clusters.

Two main density suppression factors arise in moving from clusters
to groups. These are best discussed on referring to ICM in
hydrostatic equilibrium within the DM potential well $\Delta \phi$
(Cavaliere \& Fusco-Femiano 1976; Jones \& Forman 1984). In the
corresponding ICM density run $n(r)=n_2\, \exp{[\beta\, \Delta
\phi(r)]}$ the energy injection resets the values of $\beta=T_v/T$
and of $n_2$.

One factor is related to the outflow and is given by $\beta$; the
latter is lowered by about $0.6$ from rich clusters toward poor
groups where stellar preheating provides a contribution to $T$
comparable to $T_v$. A second factor is the differential
containment expressed by the boundary value $n_2$. If this is set
by jump conditions across the accretion shocks at $r\approx R_v$
(see Takizawa \& Mineshige 1998; Gheller, Pantano \& Moscardini
1998), the density is further suppressed from clusters to groups
by another factor approaching $1/2$ on average, see Menci \&
Cavaliere (2000).

These authors model the process on grafting the ICM equilibrium
onto the semi-analytic treatment (SAM) of star and galaxy
formation. This is based on the merging histories of the DM, and
includes star formation and gas heating/ejection by SNe in terms
of simple recipes. The latter imply heating to dominate at scales
of bright galaxies and larger; only in small galaxies the gas
fractions blown out exceed $10^{-1}$ (see Madau, Ferrara \& Rees
2001).

With SN feedback, the SAMs provide detailed fits to the stellar
observables. But the agreement with the available $L_X-T$ data for
poor groups is marginal if both the Navarro, Frenk \& White (1997)
potential $\Delta \phi$ and the standard flat $\Lambda$-cosmology
are adopted. This is illustrated in next \S~3, see also Borgani et
al. (2001).

Note that the SAMs include the amount of cooling suitable for
baryon condensation to stars, which helps in lowering the ICM
density. A leading role of cooling (combined with suitable
feedback) to remove the low entropy gas has been discussed by
Bryan (2000), Muanwong et al. (2002), and Voit et al. (2002). We
will concentrate on a maximal but realistic feedback process.
%
%%%%%%%%%%%%%%%%%%%%%%%%%%%%%%%%%%%%%%%%%%%%%%%%%%%%%%%%%%%%%%%%%%%%%%%%%
%%%%%%%%%%%%%%%%%%%%%%%%%%%%%%%%%%%%%%%%%%%%%%%%%%%%%%%%%%%%%%%%%%%%%%%%%
\section{Feedback from AGNs}
%%%%%%%%%%%%%%%%%%%%%%%%%%%%%%%%%%%%%%%%%%%%%%%%%%%%%%%%%%%%%%%%%%%%%%%%%
%%%%%%%%%%%%%%%%%%%%%%%%%%%%%%%%%%%%%%%%%%%%%%%%%%%%%%%%%%%%%%%%%%%%%%%%%
%
The natural sources of strong feedback are the AGNs, energized by
accretion of cool baryons onto supermassive BHs in galactic cores
(see Wu, Fabian \& Nulsen 2000; Bower et al. 2001). The outputs
are large, of order $2\, 10^{62}\, (M_{\bullet}/10^9\, M_{\odot})$
ergs for an accreted mass $M_{\bullet}$, with the standard
mass-energy conversion efficiency of order $10^{-1}$. If a
fraction $f$ is coupled to the surrounding medium, the energy
actually injected comes to $\Delta E\approx f\, 10^{50}$ erg per
solar mass condensed into stars; we have used $2\, 10^{-3}$ for
the ratio of the BH mass to that of the current host bulges
(Ferrarese \& Merritt 2000; Gebhardt et al. 2000; see also Fabian
\& Iwasawa 1999).

Compared with SNe, AGNs potentially provide a larger energy output
in shorter times, close to $t_{d}$ of the host galaxies. However,
$f$ is more uncertain than the analogous quantity for SNe, though
is expected to be lower.

The $10\%$ radio-loud AGNs directly produce large kinetic energies
in the form of jets, but to now the observations indicate limited
impact on the gas surrounding a number of active sources (see
McNamara et al. 2001; Terashima \& Wilson 2001; Young, Wilson \&
Mundell 2002). In the $90\%$ radio-quiet AGNs a smaller coupling
$f\approx 10^{-2}$ is expected in view of their flat spectra and
of the low photon momenta. Values up to $f\sim 10^{-1}$ are only
conceivable in systems where the photons are heavily
scattered/absorbed within the gravitational reach of the BH, and
escape in hard X-rays if at all (Fabian, Wilman \& Crawford 2002).

Whence the interest in probing $f$ from overall effects on the
ICM. During the AGN activity the gas initially contained in a
large galaxy or a group will be heated up and partially blown out.
We will treat blowout and outflow driven {\it internally} in poor
groups with $kT_v\sim 1$ keV where $\Delta E/E\sim 1$. These
effects are maximized in spherical symmetry, not unfit for the
radio-quiet AGNs we consider.

Within the structure's dynamical time $t_d$ we describe the
transient regime as a blastwave sweeping through the surrounding
gas (see Platania et al. 2002); when $\Delta E/E\la 1$ holds the
leading shock at $r=R_s$ is not necessarily strong, and DM gravity
is important.

%%%%%%%%%%%%%%%%%%%%%%%%%%%%%%%%%%%%%%%%%%%%%%%%%%%%%%%%%%%%%%%%%%%%%%%%%
%                       FIG. 1
%%%%%%%%%%%%%%%%%%%%%%%%%%%%%%%%%%%%%%%%%%%%%%%%%%%%%%%%%%%%%%%%%%%%%%%%%
\vspace{0.5cm}
\centerline{{\vbox{\epsfxsize=7.5cm\epsfbox{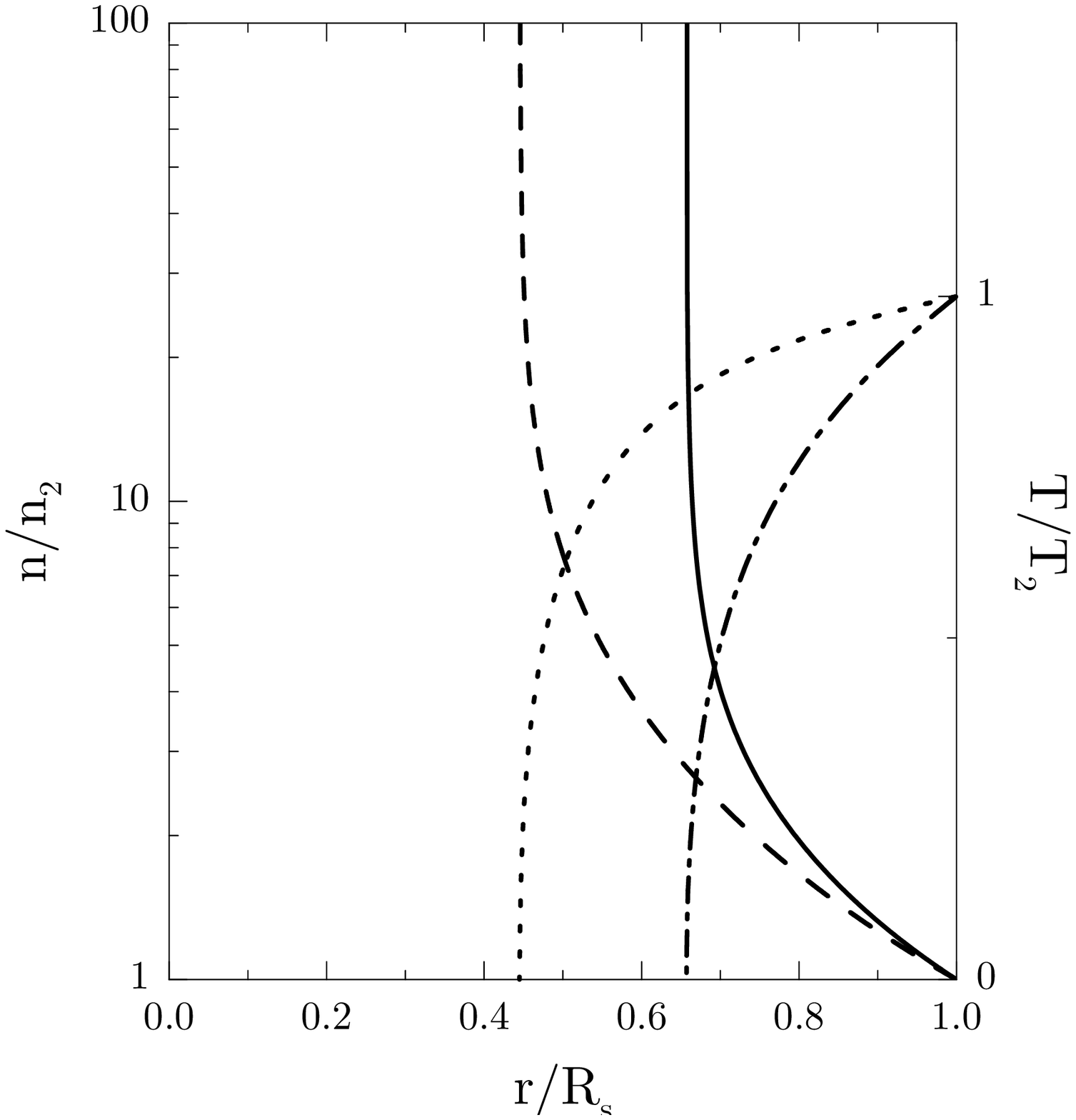}}}}
\figcaption{Radial distributions of density and temperature
(normalized to their postshock values) in a blastwave driven by a
flaring AGN. Dot-dashed and dotted lines refer to the temperature
behind a strong ($\Delta E/E= 1.8$) or a weak shock ($\Delta E/E=
0.3$), respectively; solid and dashed lines refer to the density.
On approaching the piston ($r=0.66\, R_s$ or $r=0.45\, R_s$ for
the adopted values of $\Delta E/E$) the density diverges weakly
while the mass inside $r$ vanishes.} \vspace{0.5cm}
%%%%%%%%%%%%%%%%%%%%%%%%%%%%%%%%%%%%%%%%%%%%%%%%%%%%%%%%%%%%%%%%%%%%%%%%%
%
From the relevant hydrodynamical equations we have derived (and
illustrate in fig.~1) a family of self-similar solutions of the
Sedov (1959) type that include the DM gravity, a steep initial
density gradient, and centrally injected energy $\Delta E(t)$
growing over times of order $t_{d}$. The radiative cooling is slow
on mass-average in our groups.

Our fiducial case will have $\Delta E(t)\propto t$ injected into
an initial configuration with  $n(r)\propto r^{-2}$, i.e.,
isothermal ICM in hydrostatic equilibrium ($\beta\approx 1$) in
the potential provided by DM density $\rho(r)\propto r^{-2}$; we
denote by $E(R_s)$ the modulus of the total initial energy within
the shock radius $R_s$. Then the leading shock moves outward at a
costant speed $v_s$, only moderately supersonic when $\Delta
E/E\la 1$.

Self-similarity implies $\Delta E(t)/E(R_s)$ to be independent of
time and position, as is especially simple to see in our fiducial
case where $E(R_s)\propto R_s\propto t\propto \Delta E(t)$. For
two values of $\Delta E/E$ we show in fig.~1 the density and
temperature runs. The flow begins at a `piston', the inner contact
surface where the density diverges weakly while the gas mass
within $r$ and the temperature $T(r)$ vanish.

In fact, the perturbed gas is confined to a shell with inner
(piston) radius $\lambda\, R_s$ and outer (shock) radius $R_s$.
Self-similarity implies the thickness $\Delta R_s/R_s = 1-\lambda$
of such a shell to depend only on $\Delta E/E$; for strong shocks
driven by $\Delta E/E\gg 1$ we find $\lambda\rightarrow 0.84$,
while for a weak shock corresponding to $\Delta E/E = 0.3$ we find
$\lambda = 0.45$.

These considerations lead us to represent our solutions in terms
of the simple `shell approximation', known to provide results
reliable to better than $15\%$, see Cavaliere \& Messina (1976);
Ostriker \& McKee (1988). In this approximation the energy balance
reads
\smallskip
\begin{equation}
\Delta E + E = {1\over 2}\, m\, v_2^2 + {3\over 2}\, \bar{p}\,
V-{G\, M\, m\over R_s}~,
\end{equation}
and directly shows the relevance of $\Delta E/E$. Here $M$ is the
DM mass within $R_s$; also, $V = 4\pi R_s^3\, (1-\lambda^3)/3$ is
the volume of the shell; $m$ and $\bar{p}$ are the associated gas
mass and mean pressure; finally, $v_2\propto v_s$ is the postshock
velocity given by the Rankine-Hugoniot conditions.

%
%%%%%%%%%%%%%%%%%%%%%%%%%%%%%%%%%%%%%%%%%%%%%%%%%%%%%%%%%%%%%%%%%%%%%%%%%
%                           FIG. 2
%%%%%%%%%%%%%%%%%%%%%%%%%%%%%%%%%%%%%%%%%%%%%%%%%%%%%%%%%%%%%%%%%%%%%%%%%
\vspace{0.5cm}
\centerline{{\vbox{\epsfxsize=7.5cm\epsfbox{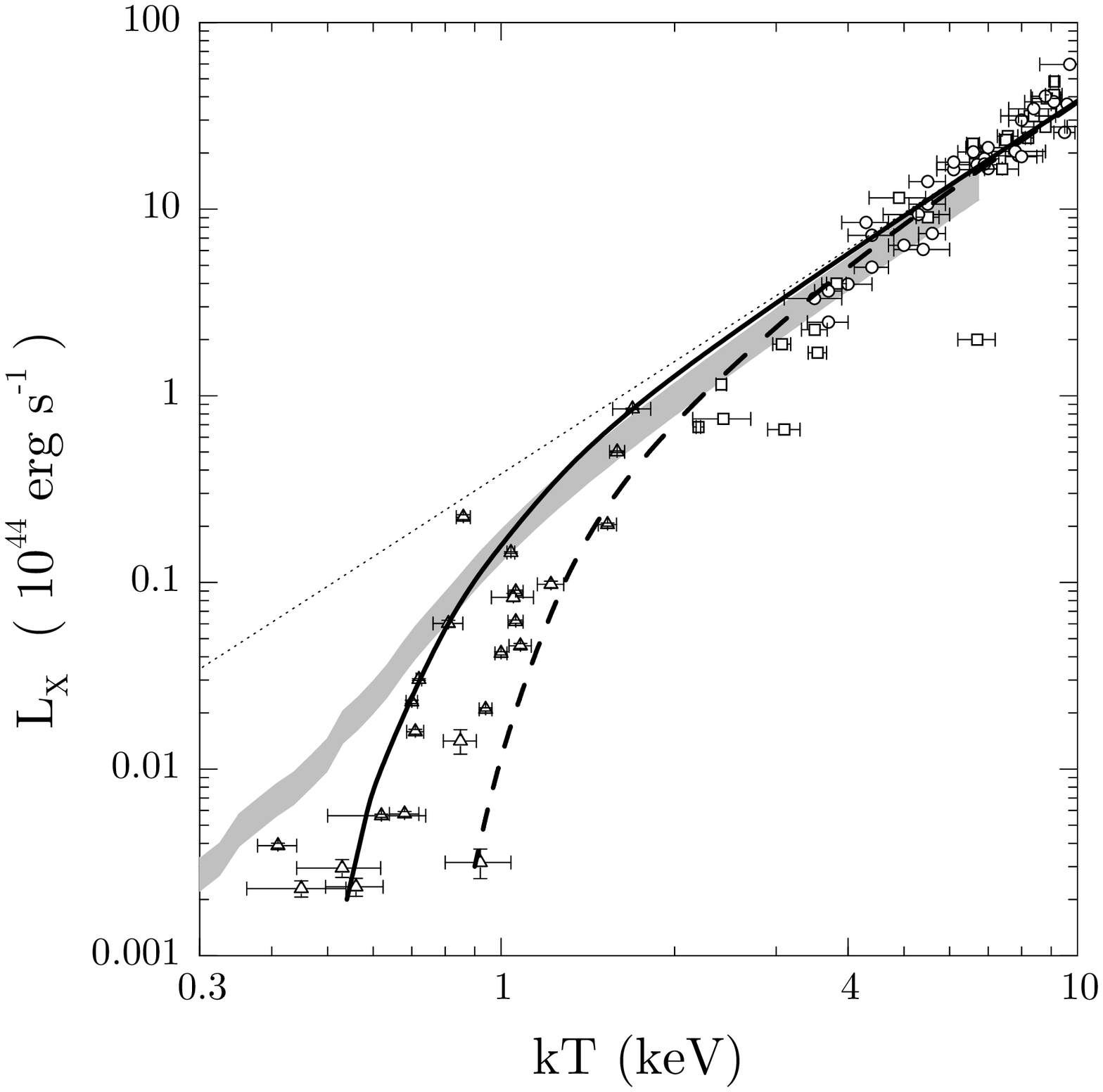}}}}
\figcaption{The $L_X-T$ correlation; bolometric $L_X$ including
standard line emissions. Thin dotted line: gravitational scaling
$L\propto T_v^2$. The shaded one-sigma strip results from a SAM
including the stochastic merging histories of the DM and the SN
feedback, see \S~2. Thick lines: our results with feedback from
AGNs accreting as specified at the end of \S~3, and with $f = 3\,
10^{-2}$ (solid) and $f = 10^{-1}$ (dashed). Data: Markevitch
(1998, circles), Arnaud \& Evrard (1999, squares), Helsdon \&
Ponman (2000, triangles).} \vspace{0.5cm}
%%%%%%%%%%%%%%%%%%%%%%%%%%%%%%%%%%%%%%%%%%%%%%%%%%%%%%%%%%%%%%%%%%%%%%%%%
%

Self-similarity requires all terms in eq.~(1) to scale like $R_s$;
the coefficients depend only on $\Delta E/E$, and are easily
derived following the pattern indicated by the above authors. We
find the ratio of the kinetic to the thermal energy (i.e., the
$1^{\mathrm{st}}$ to the $2^{\mathrm{nd}}$ term on the r.h.s. of
eq.~1) to range from $5\, 10^{-2}$ up to $2$ when $\Delta E/E$
increases from $0.3$ to values larger than $1$. Analogously, one
may derive the dependence of $v_s$ on $\Delta E/E$, and an
analytic approximation to the mass distribution within the shell.

After the passage of the blastwave, and before a major merging
event reshuffles the DM mass substantially, the gas recovers
hydrostatic equilibrium and again $n(r)=n_2\, \exp{[\beta\, \Delta
\phi]}$ holds; but now the governing parameters are those given in
Table~1. The value of $\beta=T_v/\bar{T}$ (related to outflow) is
reset using the mass-averaged temperature $\bar{T}$. The new ICM
mass $m-\Delta m$ (left behind by the blowout) is taken to be that
still residing at $t=t_d$ between the piston and $R_v$; thus the
boundary condition $n_2$ is reset by requiring consistency with
the volume integral of $n(r)$.

We then compute $L_X\propto\bar{T}^{1/2}\, \int{dr\, r^2\,
n^2(r)}$ and plot it as a function of temperature in fig.~2; here
we approximate $\bar{T}$ with $T_v$, since these differ only
modestly as seen from the values of $\beta$ in Table~1. Our
results are given for two values of the energy coupled; these
bracket $\Delta E = 2\, 10^{60}$ erg corresponding, e.g., to
$f\approx 5\, 10^{-2}$ and to $M_{\bullet} = 10^9\, M_{\odot}$ for
the largest BH (or sum of BHs) formed within $t_d$ in groups with
membership around $10$ bright galaxies.

The variable $T_v$ may be related to the quantity $\Delta E/E$
that governs the blastwave on using $E\propto$ $(kT_v)^{5/2}$, see
\S~1. Assuming $\Delta E\propto M_{\bullet}$ (see the beginning of
this Sect.), the correspondence is simply given by
\begin{equation}
{\Delta E\over E} = 0.1\, \left({f\over 10^{-2}}\right)\,
\left({M_{\bullet}\over 10^9\, M_{\odot}}\right)\,
\left({kT_v\over \mathrm{keV}}\right)^{-5/2}~.
\end{equation}
In fig.~2 we have actually implemented the second approximation
$\Delta E\propto M_{\bullet}\, [1+(kT_v/\mathrm{keV})^{6/7}]/2$.

This accounts for the cosmological evolution of the AGNs that
occurs at redshifts $z < 2.5$, when groups and clusters with
increasing $kT_v\propto M^{2/3}_v\, (1+z)$ are formed by the
standard hierarchical clustering. In the critical universe the
above (normalized) correction stems from: (a) the diminishing
fraction of member galaxies activated within the increasing $t_d$,
that scales simply as $M_v$, i.e., approximately as $(1+z)^{-5}$;
(b) all outputs weakening as $(1+z)^3$ (see Cavaliere \& Vittorini
2002). In the standard flat $\Lambda$-cosmology the result
provides a close upper limit.

In fig.~2 we also recall the contribution from SNe, and report the
available X-ray data.
%
%%%%%%%%%%%%%%%%%%%%%%%%%%%%%%%%%%%%%%%%%%%%%%%%%%%%%%%%%%%%%%%%%%%%%%%%%
%                           TABLE 1
%%%%%%%%%%%%%%%%%%%%%%%%%%%%%%%%%%%%%%%%%%%%%%%%%%%%%%%%%%%%%%%%%%%%%%%%%
\tabcaption{Parameters of the recovered equilibrium}
\begin{center}
\begin{tabular}{cccccc}
\tableline\tableline\noalign{\smallskip}
$\Delta E/E$ & $\beta$ & $1-\Delta m /m$\\
\noalign{\smallskip} \tableline\noalign{\smallskip}
0.3 &0.94  &0.92  \\
1   &0.86  &0.58  \\
1.8   &0.8   &0.16 \\
\noalign{\smallskip} \tableline \noalign{\smallskip}
\end{tabular}

\smallskip
{\small $^{\star}$ The approximation $\Delta m/m\approx 0.5\,
\Delta E/E$ holds to better than $10\%$ for $\Delta E/E < 1.4$.}
\end{center}
%%%%%%%%%%%%%%%%%%%%%%%%%%%%%%%%%%%%%%%%%%%%%%%%%%%%%%%%%%%%%%%%%%%%%%%%%
%
%%%%%%%%%%%%%%%%%%%%%%%%%%%%%%%%%%%%%%%%%%%%%%%%%%%%%%%%%%%%%%%%%%%%%%%%%
%%%%%%%%%%%%%%%%%%%%%%%%%%%%%%%%%%%%%%%%%%%%%%%%%%%%%%%%%%%%%%%%%%%%%%%%%
\section{Discussion and Conclusions}
%%%%%%%%%%%%%%%%%%%%%%%%%%%%%%%%%%%%%%%%%%%%%%%%%%%%%%%%%%%%%%%%%%%%%%%%%
%%%%%%%%%%%%%%%%%%%%%%%%%%%%%%%%%%%%%%%%%%%%%%%%%%%%%%%%%%%%%%%%%%%%%%%%%
%
Two key features are apparent in fig.~2 and are spelled out in
Table~1. First, into the cluster range the deviations from the
gravitational scaling vanish because both $1-\Delta m/m$ and
$\beta=T_v/\bar{T}$ saturate to $1$. Second, moving into the group
range the luminosity is non-linearly suppressed as $L_X\propto
n^2_2\propto (1-\Delta m/m)^2$, due to the increasing contribution
from the blowout as $\Delta E/E$ raises toward $1$.

The current X-ray data in groups are seen to require values around
$f\approx 5\, 10^{-2}$ in our blastwave model. With these, the
feedback from AGNs dominates over SNe in poor groups, causing
stronger suppression of $L_X$ and further bending of the $L_X-T$
relation. Variance of $f$ from $3\, 10^{-2}$ to $10^{-1}$ produces
a widening strip, but one still consistent with the current data
and their scatter. This also covers the effects of moderate
non-spherical deviations.

As our $L_X-T$ relation intrinsically steepens towards poor
groups, we can check it in the adjoining galactic range where
cooling still does not dominate. In terms of the velocity
dispersion $\sigma  = (kT_v/0.6\, m_p)^{1/2}$ from the virial
relation, we find a steepening correlation $L_X\propto \sigma^n$;
for the upper values of $f$ the slope is $n\approx 8.5\div 10$ in
large galaxies with $\sigma=300$ km s$^{-1}$, which accords with
the detections and the fit by Mahdavi \& Geller (2001).

Our result is due to a blastwave with $\Delta E/E\approx 1.2$
causing $\Delta m/m\approx 0.6$. Down to what scales can we extend
the increasing trend of $\Delta E/E$? We argue $\Delta E/E$ {\it
on average} will not exceed $1$ by much, nor will $\Delta m/m$
attain $1$.

First, let us impose the limiting constraint $\Delta E/E\approx 1$
to eq.~(2), and find the accreted masses
\begin{equation}
M_{\bullet} \approx 2\, 10^9 M_{\odot}\, \left({f\over
10^{-2}}\right)^{-1}\, \left({\sigma\over 300\, \mathrm{km\,
s}^{-1}}\right)^5~.
\end{equation}
Such values, consistent with those adopted in our computations,
for $f$ in the range $3 - 5\, 10^{-2}$ gratifyingly agree with the
masses of dark objects detected at the center of many galaxies.
Also the trend accords with the similar correlation pointed out by
Ferrarese \& Merritt (2001) and Gebhardt et al. (2001). Note that
on using the second approximation to $M_{\bullet}$ discussed after
eq.~(2), our correlation is somewhat flatter than
$M_{\bullet}\propto \sigma^{5}$; it has the slope $4.3$ around
$\sigma\approx 300$ km s$^{-1}$, and the prefactor $3\, 10^9\,
M_{\odot}$.

Next we discuss how our limiting value $\Delta E/E\approx 1$
arises in galaxies from BH accretion regulated by the AGN itself
(see also Silk \& Rees 1998). On the one hand, sustaining $\Delta
E/E$ to about $1$ requires sufficient cold gas made available for
inflow. The requirement is met by gravitational torques exerted in
the host by companion galaxies within small groups, see Cavaliere
\& Vittorini (2002). During encounter or flyby times of order
$t_d$ such interactions destabilize fractional gas masses of order
$10^{-2}$, while the values needed to satisfy eq.~(3) in the host
galaxies are only of order $\sigma^2/f\, \eta\, c^2\approx 2\,
10^{-3}\, (\sigma/300 \,\mathrm{km\, s^{-1}})^2$.

On the other hand, $\Delta E/E$ will be limited if the accretion
itself can be affected on the time scale $t_d$ by the AGN
feedback; if so, the AGN will fade out. Declining luminosities are
included in our self-similar blastwave family under the form $L
\propto t^{5\,(2-\omega)/\omega}$ if the initial density gradient
follows a steeper law $n\propto r^{-\omega}$ with $\omega \geq 2$.
Increasing $\omega$ up to $2.5$ corresponds to $L(t)$ going from
constant to a spike; up to $\omega\approx 2.4$ the non-linear
behavior of $L_X-T$ around $\Delta m/m\approx 1/2$ is generic.

But when $\omega$ approaches $2.5$ the time scales effective for
$L(t)$ become quite shorter than $t_d$; the corresponding
blastwaves cause larger values of $\Delta m/m$ at a given $\Delta
E/E$. This behavior is indicative of runaway conditions prevailing
when a galaxy happens to grow a large BH in short times. Then most
galactic gas is blown away beyond $R_v$, so the star formation
activity is suppressed already at $z\approx 2$ (see also Granato
et al. 2001). Such may have been the case for some of the recently
discovered EROs (see Cimatti et al. 2002; Alexander et al. 2002).

To sum up, we find that the AGN feedback sharply {\it steepens}
the $L_X-T$ correlation in the poor group range, and {\it links}
its shape to that of the galactic $M_{\bullet}-\sigma$
correlation. This is because in moving from clusters to groups the
energy $\Delta E$ injected by AGNs within $t_d$ grows relative to
the unperturbed $E$, and {\it overwhelms} SNe. But on entering the
galactic range, $\Delta E/E$ approaches unity and constrains the
accretion itself. These correlations provide two linked but
observationally independent {\it probes} of the hidden parameter
$f$. On the basis of our blastwave model, the existing X-ray data
concur with the optical ones to indicate values around $f\approx
5\, 10^{-2}$.
%
%%%%%%%%%%%%%%%%%%%%%%%%%%%%%%%%%%%%%%%%%%%%%%%%%%%%%%%%%%%%%%%
%%%%%%%%%%%%%%%%%%%%%%%%%%%%%%%%%%%%%%%%%%%%%%%%%%%%%%%%%%%%%%%
\acknowledgements
%%%%%%%%%%%%%%%%%%%%%%%%%%%%%%%%%%%%%%%%%%%%%%%%%%%%%%%%%%%%%%%
%%%%%%%%%%%%%%%%%%%%%%%%%%%%%%%%%%%%%%%%%%%%%%%%%%%%%%%%%%%%%%%
We thank G. De Zotti for stimulating discussions and our referee
for constructive comments.
%%%%%%%%%%%%%%%%%%%%%%%%%%%%%%%%%%%%%%%%%%%%%%%%%%%%%%%%%%%%%%%
%%%%%%%%%%%%%%%%%%%%%%%%%%%%%%%%%%%%%%%%%%%%%%%%%%%%%%%%%%%%%%%
%
%%%%%%%%%%%%%%%%%%%%%%%%%%%%%%%%%%%%%%%%%%%%%%%%%%%%%%%%%%%%%%%
%%%%%%%%%%%%%%%%%%%%%%%%%%%%%%%%%%%%%%%%%%%%%%%%%%%%%%%%%%%%%%%

%%%%%%%%%%%%%%%%%%%%%%%%%%%%%%%%%%%%%%%%%%%%%%%%%%%%%%%%%%%%%%%
%%%%%%%%%%%%%%%%%%%%%%%%%%%%%%%%%%%%%%%%%%%%%%%%%%%%%%%%%%%%%%%
%
\end{document}